  \PassOptionsToPackage{unicode}{hyperref}
  \PassOptionsToPackage{hyphens}{url}
  \PassOptionsToPackage{dvipsnames,svgnames,x11names}{xcolor}
  \documentclass[10pt,a4paper,onecolumn]{article}
  \usepackage{marginnote}
  \usepackage[rgb,svgnames]{xcolor}
  \usepackage{authblk,etoolbox}
  \usepackage{titlesec}
  \usepackage{calc}
  \usepackage{tikz}
  \usepackage[pdfa]{hyperref}
  \usepackage{hyperxmp}
  \immediate\pdfobj stream attr{/N 3} file{sRGB.icc}
  \pdfcatalog{%
    /OutputIntents [
      <<
        /Type /OutputIntent
        /S /GTS_PDFA1
        /DestOutputProfile \the\pdflastobj\space 0 R
        /OutputConditionIdentifier (sRGB)
        /Info (sRGB)
      >>
    ]
  }
  \usepackage{caption}
  \usepackage{orcidlink}
  \usepackage{tcolorbox}
  \usepackage{amssymb,amsmath}
  \usepackage{ifxetex,ifluatex}
  \usepackage{seqsplit}
  \usepackage{xstring}

  \makeatletter
  \@ifundefined{KOMAClassName}{% if non-KOMA class
    \IfFileExists{parskip.sty}{%
      \usepackage{parskip}
    }{% else
      \setlength{\parindent}{0pt}
      \setlength{\parskip}{6pt plus 2pt minus 1pt}}
  }{% if KOMA class
    \KOMAoptions{parskip=half}}
  \makeatother
  \usepackage{longtable,booktabs,array}
  \usepackage{calc} % for calculating minipage widths
  \usepackage{etoolbox}
  \makeatletter
  \patchcmd\longtable{\par}{\if@noskipsec\mbox{}\fi\par}{}{}
  \makeatother
  \IfFileExists{footnotehyper.sty}{\usepackage{footnotehyper}}{\usepackage{footnote}}
  \makesavenoteenv{longtable}
  \NewDocumentCommand\citeproctext{}{}
  \NewDocumentCommand\citeproc{mm}{%
    \begingroup\def\citeproctext{#2}\cite{#1}\endgroup}
  \makeatletter
   \let\@cite@ofmt\@firstofone
   \def\@biblabel#1{}
   \def\@cite#1#2{{#1\if@tempswa , #2\fi}}
  \makeatother
  \newlength{\cslhangindent}
  \newlength{\csllabelwidth}
  \newenvironment{CSLReferences}[2] % #1 hanging-indent, #2 entry-spacing
   {\begin{list}{}{%
    \setlength{\itemindent}{0pt}
    \setlength{\leftmargin}{0pt}
    \setlength{\parsep}{0pt}
    \ifodd #1
     \setlength{\leftmargin}{\cslhangindent}
     \setlength{\itemindent}{-1\cslhangindent}
    \fi
    \setlength{\itemsep}{#2\baselineskip}}}
   {\end{list}}
  \usepackage{calc}

  \ifLuaTeX
  \usepackage[bidi=basic,shorthands=off,]{babel}
  \else
  \usepackage[bidi=default,shorthands=off,]{babel}
  \fi
  \ifLuaTeX
    \usepackage{selnolig} % disable illegal ligatures
  \fi
  \usepackage{float}
  \let\origfigure\figure
  \let\endorigfigure\endfigure
  \renewenvironment{figure}[1][2] {
      \expandafter\origfigure\expandafter[H]
  } {
      \endorigfigure
  }

  \usepackage[top=3.5cm, bottom=3cm, right=1.5cm, left=1.0cm,
              headheight=2.2cm, reversemp, includemp, marginparwidth=4.5cm]{geometry}

  \titleformat{\section}
    {\normalfont\sffamily\Large\bfseries}
    {}{0pt}{}
  \titleformat{\subsection}
    {\normalfont\sffamily\large\bfseries}
    {}{0pt}{}
  \titleformat{\subsubsection}
    {\normalfont\sffamily\bfseries}
    {}{0pt}{}
  \titleformat*{\paragraph}
    {\sffamily\normalsize}

  \usepackage{fancyhdr}
  \makeatletter
  \let\ps@plain\ps@fancy
  \fancyheadoffset[L]{4.5cm}
  \fancyfootoffset[L]{4.5cm}

  \definecolor{linky}{rgb}{0.0, 0.5, 1.0}

  \ifPDFTeX\else
  \fi
  \IfFileExists{upquote.sty}{\usepackage{upquote}}{}
  \IfFileExists{microtype.sty}{% use microtype if available
    \usepackage[]{microtype}
    \UseMicrotypeSet[protrusion]{basicmath} % disable protrusion for tt fonts
  }{}

  \PassOptionsToPackage{usenames,dvipsnames}{color} % color is loaded by hyperref
  \usepackage{lineno}

  \ifLuaTeX
    \usepackage{selnolig}  % disable illegal ligatures
  \fi

  \title{COPASI Skills: Agent Skills for COPASI analyses through Basico}

  \author[1]{Frank T. Bergmann%
    \,\orcidlink{0000-0001-5553-4702}\,
  }
  \author[1]{Sven Sahle%
    \,\orcidlink{0000-0002-5458-7404}\,
  }

  \affil[1]{BioQUANT / Heidelberg University, INF 267, Heidelberg, Germany  }
  
\begin{document}
  \maketitle

  \marginpar{

    \begin{flushleft}
    %\hrule
    \sffamily\small

    \end{flushleft}
  }

  \section{Summary}\label{summary}

  Biochemical models represent biological systems as networks of reactions
  that describe the production, consumption, and transformation of
  molecular species over time. These models provide a quantitative
  framework for analyzing system dynamics and enable computational
  experiments that complement traditional wet-lab approaches, helping
  researchers gain mechanistic insight and make predictive assessments of
  biological behavior. COPASI is a widely used program for simulating and
  analyzing those models. These tasks are made availabe to Python through
  Basico. COPASI Skills is a set of instruction files that tell an
  assistant how to drive Basico for one analysis at a time. Each skill is
  a directory containing a \texttt{SKILL.md} file. A short description at
  the top tells the assistant when the skill applies, and the rest of the
  file states which calls to make, which files to write, and which
  potential neighboring skill to choose additionally. A researcher can ask,
  for example, in ordinary language, for a time course, a fit, or a
  published model. The assistant follows the matching skill and leaves
  tables or a report next to the model. The collection is released under a
  two-clause BSD license from
  \url{https://github.com/fbergmann/copasi-skills}.

  \section{Statement of need}\label{statement-of-need}

  The COmplex PAthway SImulator {[}\texttt{COPASI}; Hoops et al.
  (\citeproc{ref-hoops2006copasi}{2006}); Bergmann et al.
  (\citeproc{ref-bergmann_2017}{2017}){]} simulates deterministic and
  stochastic time courses, computes steady states, and runs optimization,
  parameter estimation, sensitivity analysis, and metabolic control
  analysis, using SBML (\citeproc{ref-hucka2003sbml}{Hucka et al., 2003})
  or its own file format. Basico
  (\citeproc{ref-bergmann2023basico}{Bergmann, 2023}) exposes those tasks
  as Python functions and is the interface a script should call. But an
  assistant placed in front of COPASI / Basico does not, by itself, know
  the workflow a modeler expects. This project contributes the procedures
  that sit on top of both, so that an assistant will follow the same
  approach in a reliable way.

  That procedure now has a shared file format. In October 2025, Anthropic
  described Agent Skills as directories whose \texttt{SKILL.md} file
  begins with a name and a description
  (\citeproc{ref-anthropic2025skills}{Anthropic, 2025a}). The assistant
  loads those two fields for every installed skill and loads the
  instructions only when a request matches the description. Anthropic
  argues that the same pattern suits scientific protocols, because a skill
  can make an assistant follow a stated procedure, and it demonstrated a
  skill for quality control of single-cell RNA sequencing data
  (\citeproc{ref-anthropic2025lifesciences}{Anthropic, 2025b}). The format
  is now an open specification. The specification requires the description
  to say both what the skill does and when to use it, and it treats
  further files as material to be read only when the instructions ask for
  them (\citeproc{ref-agentskills2026}{Agent Skills, 2026}). Scientific
  skill collections have since packaged that layout for other research
  software and databases (\citeproc{ref-kdense2026skills}{K-Dense, 2026}).
  However, it does not encode the functionality COPASI provides.

  This contribution is a COPASI-specific set of skills encoded in the
  Agent Skills format. Each description names the question it answers and
  potentially when not to use it, a request for control coefficients is
  not treated as a parameter sensitivity, and a model with no steady state
  is handed to a diagnostic skill.

  Anyone can write a \texttt{SKILL.md} file that drives COPASI. This
  collection is published and reviewed by the COPASI authors rather than
  from an unreviewed prompt. We invite further skills, and invite
  contributions whenever the present skills fail, so that the procedures
  can be improved.

  \section{Software design}\label{software-design}

  The main design choice is one skill per analysis. A single skill that
  described every COPASI task would match almost any modeling question and
  would load a long text whenever it matched. Separate descriptions stay
  specific. That matters because the description is the only text the
  assistant holds for every installed skill, while the body is read after
  a match (\citeproc{ref-agentskills2026}{Agent Skills, 2026};
  \citeproc{ref-anthropic2025skills}{Anthropic, 2025a}). The body then
  states the Basico calls, the output files, and the cases that belong to
  a different skill.

  \textbf{Note:} As the skills install packages from the Python Package
  Index, and they create and run Python files, they should be used inside
  a sandbox.

  \autoref{tab:skills} lists the seventeen skills in the collection.
  Together they cover the path from an empty environment and a published
  model through simulation and steady state, then through scans,
  sensitivities, control analysis, optimization, and parameter estimation,
  and on to structural and stochastic summaries. The overview skill is the
  one that reads a model and the tasks already configured inside it and
  writes a linked report, rather than answering a single numerical
  question.

  Whenever steps in a procedure could easily lead to mistakes, the skill
  directory contains a Python script and \texttt{SKILL.md} tells the
  assistant when to run it. Examples of this are the diagnosis of a
  missing steady state or the evaluation of a finished parameter
  estimation.

  Results are CSV tables or, for the model overview, a linked Markdown
  report that can be archived with the model. The assistant keeps every
  skill description, opens one \texttt{SKILL.md} when a description
  matches, and follows it into Basico, which calls COPASI and writes the
  files. The same skills can be installed for more than one assistant,
  which is why the project does not add a private file format or a second
  Python API.

  \begin{longtable}[]{@{}
    >{\raggedright\arraybackslash}p{(\linewidth - 2\tabcolsep) * \real{0.5000}}
    >{\raggedright\arraybackslash}p{(\linewidth - 2\tabcolsep) * \real{0.5000}}@{}}
  \caption{The seventeen skills and an example question each one answers.
  \label{tab:skills}}\tabularnewline
  \toprule\noalign{}
  \begin{minipage}[b]{\linewidth}\raggedright
  Skill
  \end{minipage} & \begin{minipage}[b]{\linewidth}\raggedright
  Question it answers
  \end{minipage} \\
  \midrule\noalign{}
  \endfirsthead
  \toprule\noalign{}
  \begin{minipage}[b]{\linewidth}\raggedright
  Skill
  \end{minipage} & \begin{minipage}[b]{\linewidth}\raggedright
  Question it answers
  \end{minipage} \\
  \midrule\noalign{}
  \endhead
  \bottomrule\noalign{}
  \endlastfoot
  \texttt{copasi-setup} & Is the Python environment able to import
  Basico? \\
  \texttt{copasi-model-search} & Which published model should be
  loaded? \\
  \texttt{copasi-model-edit} & How should the model be modified for the
  next analysis? \\
  \texttt{copasi-model-values} & How should model values be accessed or
  updated? \\
  \texttt{copasi-timecourse} & How do the variables move over time? \\
  \texttt{copasi-steadystate} & What is the steady state, if there is
  one? \\
  \texttt{copasi-reach-steadystate} & Why is there no steady state, and
  what change produces one? \\
  \texttt{copasi-scan} & How do results change across a parameter scan? \\
  \texttt{copasi-sensitivity} & Which parameters move a chosen result the
  most? \\
  \texttt{copasi-mca} & Which reactions control a flux or a
  concentration? \\
  \texttt{copasi-optimization} & Which parameter values minimize or
  maximize a quantity? \\
  \texttt{copasi-parameter-estimation} & Which parameter values agree with
  the measurements? \\
  \texttt{copasi-parameter-estimation-evaluation} & Is that fit
  identifiable? \\
  \texttt{copasi-efm} & Which minimal pathways can carry a steady-state
  flux? \\
  \texttt{copasi-lyapunov} & Do nearby trajectories diverge? \\
  \texttt{copasi-lna} & What is the linear-noise covariance around a
  steady state? \\
  \texttt{copasi-model-overview} & What is in this model, and what do its
  configured tasks report? \\
  \end{longtable}

  \section{Target audience}\label{target-audience}

  This project is intended for systems biologists, students and
  instructors using computational models encodable in SBML or COPASI
  models and who want a repeatable analysis rather than a one-off chat.

  The results still have to be checked. COPASI Skills is there to help a
  modeler’s own analysis and to offer suggestions along the way. It is not
  there so that the thinking can be handed to the assistant. The person
  who asked for the analysis remains responsible for deciding whether a
  table, a fit, or a suggested change is meaningful.

  Of course, the other target is the assistant. The instructions do not
  depend on one vendor’s tool protocol. They simply say which Basico
  functions to run and which files to write, so any host (like GitHub
  Copilot, Claude Code, Cursor) that reads the Agent Skills format can
  load these skills.

  \section{AI usage disclosure}\label{ai-usage-disclosure}

  The manuscript and the helper scripts under \texttt{skills/} were
  drafted with Cursor using Grok 4.7 and with GPT-5.6 Sol through Copilot
  Pro. The author revised both after testing them against real analyses
  and discussions with team members, and checked the skill inventory, the
  description of COPASI and Basico, and every citation. The published
  software does not call a language model when it runs.

  \section{Acknowledgements}\label{acknowledgements}

  FTB thanks the COPASI developers, currently Stefan Hoops, Ursula Kummer,
  Jürgen Pahle, and Sven Sahle for contributions and discussions. This
  work has been possible thanks to LiBiS, the Baden-Württemberg Institute
  for Bioinformatics Infrastructure, and by the BMBF-funded de.NBI
  initiative (031L0104A, W-de.NBI-016).

  \section*{References}\label{references}
  \addcontentsline{toc}{section}{References}

  \protect\phantomsection\label{refs}

  \end{document}